\newcommand{\abs}[1]{\left\vert#1\right\vert}

\newcommand{\ket}[1]{\left\vert #1\right\rangle}

\newcommand{\bra}[1]{\left\langle #1\right\vert}

\newcommand{\brkt}[2]{\left\langle #1 \vert #2\right\rangle}

\newcommand{\braket}[3]{\left\langle #1 \right\vert #2\left\vert #3\right\rangle}

\newcommand{\brak}[1]{\left\langle #1\right\rangle}
\newcommand{\brakt}[1]{\left\langle #1\right\rangle_{3}}
\newcommand{\braks}[1]{\left\langle #1\right\rangle_{S}}

\newcommand{\detcr}[3]{\ensuremath{  \left[\left[#1 , #2 \right] , #3 \right] }  }
\newcommand{\D}{\displaystyle}

\documentclass[preprint,aps,prc,showpacs,amsmath,amssymb]{revtex4-1}
\usepackage{times}          
\usepackage{graphicx}
\usepackage{bm}
\usepackage{amsmath}
\usepackage{amsxtra}
\usepackage{amstext}
\usepackage{amssymb}
\usepackage{latexsym}
\usepackage{dsfont}

\begin{document}
\title{{\large  magnetic moments of the ground-state $\mathbf{J^P=\frac{3}{2}}^{+}$ baryon decuplet }}
\author{Milton Dean Slaughter}
\address{Department of Physics, Florida International University, Miami, Florida 33199, USA}
\email{slaughtm@FIU.Edu,\;\;Slaughts@PhysicsResearch.Net}

\begin{abstract}
The magnetic moment---a function of the electric charge form factor $F_{1}(q^{2})$ and the magnetic dipole form factor $F_{2}(q^{2})$ at zero four-momentum transfer $q^{2}$---of the ground-state $J^{P}=\frac{3}{2}^{+}$ baryon decuplet magnetic moments have been studied for many years with limited success.  At present, only the magnetic moment of the $\Omega^{-}$ has been accurately determined.  We calculate nonperturbatively the magnetic moments of the \emph{physical baryon decuplet $J^{P}=\frac{3}{2}^{+}$} members and in particular, we obtain $\mu_{\Delta^{++}}= (+3.67 \pm 0.07) \,\mu_{N}$, $\mu_{\Delta^{+}}= (+1.83 \pm 0.04) \,\mu_{N}$, $\mu_{\Delta^{0}}= (0 ) \,\mu_{N}$, and the magnetic moments of their $U$-Spin partners in terms of $\Omega^{-}$ magnetic moment data.
\end{abstract}
\maketitle

\section{Introduction}

The properties of the ground-state $J^{P}=\frac{3}{2}^{+}$ baryon decuplet magnetic moments $\Delta$, $\Xi^{*}$, $\Sigma^{*}$ and $\Omega^{-}$ have been studied for many years with limited success.  Although the masses (pole or otherwise) and decay aspects and other physical observables of some of these particles have been ascertained, the magnetic moments of many are yet to be determined.  From the Particle Data Group \cite{Nakamura:2010zzi}, only the magnetic moment of the $\Omega^{-}$ \cite{Wallace:1995pf} has been accurately determined. The magnetic moment is a function of the electric charge form factor $F_{1}(q^{2})$ and the magnetic dipole form factor $F_{2}(q^{2})$ at zero four-momentum transfer $q^{2}\equiv -Q^{2}$. The lack of experimental data for the decuplet particle members is associated with their very short lifetimes (many available strong interaction decay channels) and the existence of nearby particles with quantum numbers that allow for configuration mixing greatly increasing the difficulty of experimental determination of physical observables. The $\Omega^{-}$ (strangeness $S=-3$) is an exception in that it is composed of three valence $s$ quarks that make its lifetime substantially longer (weak interaction decay) than any of its decuplet partners.  However, even for the $\Omega^{-}$, away from the static ($q^{2}= 0$) limit, the electric charge and magnetic dipole form factors are not known. Theoretical models abound: Beg \emph{et al.} \cite{Beg:1964nm} and Gerasimov \cite{Gerasimov:1966}, and Lichtenberg \cite{Lichtenberg:1978pc} provide
excellent sources of methodological information.

In Ref.~\cite{Slaughter:2011??}, we illustrated how one may calculate the magnetic moments of the \emph{physical decuplet $U$-Spin $=\frac{3}{2}$ quartet members} (the $\Delta^{-}$, $\Sigma^{*\,-}$, and $\Xi^{*\,-}$) in terms of that of the $\Omega^{-}$ ($U$-Spin $=\frac{3}{2}$ as well) without ascribing any specific form to their quark structure or intra-quark interactions \cite{Slaughter:2011??,Oneda:1970ny,Oneda:1985wf,Slaughter:1988hx,Oneda:1989ik,alfaro63}. Theoretical and computational investigations and reviews involving the magnetic moments of the $\Omega^{-}$ and the $\Delta^{-}$ and lattice quantum chromodynamics (LQCD) (quenched and unquenched, unphysical pion mass) techniques are also available \cite{Boinepalli:2009sq,Aubin:2009qp,Aubin:2010jc}, \cite{Pascalutsa:2006up}.

The electromagnetic current $j^{\mu}_{em}(0)$ obeys the \emph{double ETCRs} $\detcr{j^{\mu}_{em}(0)}{V_{\pi^{+}}}{V_{\pi^{-}}}$ \linebreak[4]$=$ $\detcr{j^{\mu}_{em}(0)}{A_{\pi^{+}}}{A_{\pi^{-}}}$ $=2j^{\mu}_{em\;3}(0)$ and $\detcr{j^{\mu}_{em}(0)}{V_{\pi^{+}}}{V_{\pi^{-}}}$ $=$ $\detcr{j^{\mu}_{em\;3}(0)}{V_{\pi^{+}}}{V_{\pi^{-}}}$ \cite{Oneda:1979mp}---$V_{\pi^{+}}$ and $V_{\pi^{-}}$ are vector charge generators, $A_{\pi^{+}}$ and $A_{\pi^{-}}$ are axial-vector charge generators, and $j^{\mu}_{em\;3}(0)$ is the isovector part of $j^{\mu}_{em}(0)$---even in the presence of symmetry breaking. The double ETCRs, in addition to ETCRs involving axial-vector charges \cite{gell-mann63,adler,Weisberger}, allow us to relate form factors---$F_{1}(q^{2})$ and $F_{2}(q^{2})$ where $U$-Spin \textbf{\emph{is not restricted to}} $\frac{3}{2}$---associated with the $U$-spin $=\frac{3}{2}$ $\Delta^{-}$ (and hence the  $\Sigma^{*\,-}$ and $\Xi^{*\,-}$, and the $\Omega^{-}$) with those associated with decuplet members having $U$-spin $=1$ (the $\Delta^{0}$, $\Sigma^{*\,0}$, and $\Xi^{*\,0}$), $U$-spin $=\frac{1}{2}$ (the $\Delta^{+}$, and $\Sigma^{*\,+}$), and $U$-spin $=0$ (the $\Delta^{++}$).

\section{ETCRs in the Infinite Momentum Frame and Flavor Broken Symmetry}

ETCRs involve at most one current density and involve the vector and axial-vector charge generators (the
 $V_{\alpha }$ and $A_{\alpha }$ $\{\alpha
=\pi ,K,D,F,B,\ldots .\}$) of the symmetry groups of QCD.

The physical vector charge $V_{K^{0}}$ is $V_{K^{0}}=V_{6}+iV_{7}$, the physical vector charge $V_{\pi^{\pm}}$ is $V_{\pi^{\pm}}=V_{1}\pm iV_{2}$, and the physical electromagnetic current $j_{em}^{\mu}(0)$ may be written ($u$, $d$%
, $s$, $c$, $b$, $t$ quark system) as $j_{em}^{\mu}(0) =V_{3}^{\mu}(0)+(1/
\sqrt{3})V_{8}^{\mu}(0)-(2/3)^{1/2}V_{15}^{\mu}(0)+(2/5)^{1/2}V_{24}^{\mu}(0)- (3/5)^{1/2}V_{35}^{\mu}(0)
+(1\sqrt{3})V_{0}^{\mu}(0)=j^{\mu}_{V}(0)+j^{\mu}_{S}(0)$, where $j^{\mu}_{V}(0)\equiv j^{\mu}_{em\;3}(0)=$ the isovector part of the electromagnetic current, $j^{\mu}_{S}(0)\equiv$ the isoscalar part of the electromagnetic current.  One may verify that the commutation relation $\left[V_{K^{0}},j^{\mu}_{em}(0)\right]=0$ and the double ETCRs \cite{Oneda:1979mp} mentioned in the Introduction hold.

\section{the electromagnetic current matrix element}

For the on-mass shell $J^{P}={3/2}^{+}$ ground-state decuplet baryon B with mass $m_{B}$, the Lorentz- covariant and gauge-invariant electromagnetic current matrix element in momentum space with four-momentum vectors $P\equiv p_{1}+p_{2},q\equiv p_{2}-p_{1}$ ($\lambda_{1}$ and $\lambda_{2}$ denote helicity) is given by:
\begin{equation}
\left\langle B{(}p_{2},\lambda _{2})\right| j_{em}^{\mu}(0)\left|
B(p_{1}{,\lambda _{1})}\right\rangle ={\frac{e}{{(2\pi )^{3}}}}\sqrt{{%
\frac{{m}_{B}^{2}}{{E_{B}^{t}E_{B}^{s}}}}}\bar{u}^{\alpha}_{B}(p_{2},\lambda _{2})\left[ {%
\Gamma^{\mu }_{\alpha \beta}}\right] u^{\beta}_{B}\left( p_{1},\lambda _{1}\right),  \label{matrixeqn}
\end{equation}
\begin{equation}
\begin{array}{ccl}
\Gamma^{\mu }_{\alpha \beta} & = & g_{\alpha \beta}  \left\{  F_{1}^{B}(q^{2})\gamma^{\mu
}+\D{\frac{F_{2}^{B}(q^{2})i\sigma^{\mu \nu}}{2 m_{B}}  q_{\nu
}}   \right\}\\
& + & \D{\frac{q_{\alpha}q_{\beta}}{2 m_{B}^{2}}} \left\{ F_{3}^{B}(q^{2})\gamma^{\mu
}+\D{\frac{F_{4}^{B}(q^{2})i\sigma^{\mu \nu}}{2 m_{B}}  q_{\nu
}}   \right\},%
\end{array}
\label{matrixcoveqn}
\end{equation}
\noindent where $e=+\sqrt{4\pi\alpha}$, $\alpha=$ the fine structure constant, the $F_{i}^{B}$ are the four $\gamma^{*}BB$ form \mbox{factors}
[$F_{1}^{B}(0)\sim \mbox{electric charge in units of $e$}$, $(F_{1}^{B}(0)+F_{2}^{B}(0))\sim $ magnetic dipole moment in units of $e/(2m_{B})$] and $\Gamma^{\mu }_{\alpha \beta}$ is written in standard form \cite{Korner:1976hv}.
The electric charge multipole \mbox{amplitude} $G_{E}^{B}(q^{2})=[F_{1}^{B}(q^{2}) (3 - 2 \eta) + \eta \{F_{2}^{B}(q^{2}) (3 - 2 \eta) -
    2 (-1 + \eta) (F_{3}^{B}(q^{2}) + \eta\,F_{4}^{B}(q^{2}) )\}]/3$ [units of $e$],
the magnetic dipole multipole amplitude $G_{M}^{B}(q^{2})=[ (5 - 4 \eta) ( F_{1}^{B}(q^{2})+F_{2}^{B}(q^{2}))    -
 4\,\eta \,(-1 + \eta) \, (F_{3}^{B}(q^{2}) + F_{4}^{B}(q^{2}))  ]/5$ [units of $e/(2m_{B})$],
  the electric quadrupole multipole \mbox{amplitude} $G_{Q}^{B}(q^{2})=F_{1}^{B}(q^{2})+F_{3}^{B}(q^{2})\, (-1 + \eta)\, +\eta \, \{F_{2}^{B}(q^{2})+F_{4}^{B}(q^{2})\, (-1 + \eta)\}$ [units of $e/m_{B}^2$], and the magnetic octupole multipole amplitude $G_{O}^{B}(q^{2})=[F_{1}^{B}(q^{2})+F_{2}^{B}(q^{2})+(-1+\eta)\{F_{3}^{B}(q^{2}) +F_{4}^{B}(q^{2})\} ]\sqrt{6}$ [units of $e/(2m_{B}^3)$] where $\eta \equiv q^{2}/(4m_{B}^2)$.   $Q_{B}=$ charge of decuplet baryon $B$ in units of $e$, $\mu_{B}$ is the magnetic moment (measured in nuclear magneton units $\mu_{N}=e/(2 m)$, $m=$ proton mass) of baryon $B$.

In Eq.~(\ref{matrixeqn}), $u_{B}^{\beta}(\nu_{B},\theta,\lambda)$ is a spin $3/2$ baryon Rarita-Schwinger \cite{Rarita:1941mf} spinor  with helicity $\lambda$, three-momentum $\vec{p}$ with angle $\theta$ referred to the $\hat{z}$-axis, energy $E_{B}^{p}$, and velocity parameter $\nu_{B}=\sinh^{-1}({\abs{\vec p \,}}/m_B)$ \cite{Slaughter:2011??}. Our conventions are those of Rose \cite{rose1957:etam}.  Physical states are normalized with $\brkt{\vec{p\,'}}{\vec{p}}=\delta^{3}%
(\vec{p\,'}\,-\vec{p})$ and Dirac spinors
are normalized by $\bar{u}^{(r)}(p)u^{(s)}(p)=\delta_{rs}$, Dirac matrices are $\left\{  \gamma^{\mu},\gamma^{\nu}\right\}
=2g^{\mu\nu}$ with $\gamma _{5}\equiv
i\gamma^{0}\gamma^{1}\gamma^{2}\gamma^{3}$, where $g^{\mu\nu}=$
Diag $(1,-1,-1,-1)$ \cite{Slaughter:2008zd}.  In addition to obeying the Dirac equation, the Rarita-Schwinger spinors satisfy the subsidiary conditions $\gamma_{\mu}u^{\mu}_{B}\left( p,\lambda\right)=p_{\mu}u^{\mu}_{B}\left( p,\lambda\right)=0$.  Associated with baryon B are the four-momentum vectors $p_1$ (three-momentum $\vec{t}$ ($\vec{t}=t_z \hat{z}$), energy $E_{B}^{t}$) and $p_2$ (three-momentum $\vec{s}$ at angle $\theta$ ($0\leq\theta < \pi/2$) with the $\hat{z}$ axis, energy $E_{B}^{s}$, with $s_{z}=r t_{z}$ and $r\, (\mbox{constant})\,\geq 1$).

\section{$\mathbf{U}$-Spin $\mathbf{1}$\;, $\mathbf{\frac{1}{2}}$\;, and $\mathbf{0}$ Decuplet Baryon Magnetic Moment Relationships}

 Previously \cite{Slaughter:2011??,Slaughter:2011ui}  (\boldmath $U${\bf{-Spin}} $\frac{3}{2}$ \unboldmath {\bf{quartet only}}), we investigated magnetic moment
relationships by utilizing the commutator  $\left[V_{K^{0}},j^{\mu}_{em}(0)\right]=0$ inserted between the baryon pairs ($\bra{\Xi^{*\,-}s^{\sigma}}$,$\ket{\Omega^{-}t^{\sigma}}$),
($\bra{\Sigma^{*\,-}s^{\sigma}}$,$\ket{\Xi^{*\,-}t^{\sigma}}$), and
($\bra{\Delta^{-}s^{\sigma}}$,$\ket{\Sigma^{*\,-}t^{\sigma}}$) where each baryon ($B=\Delta^{-}\mbox{, }\Sigma^{*^-}  \mbox{, }\Xi^{*^-} \mbox{, or }\, \Omega^{-}$) had $Q_{B}=-e$, helicity $+3/2$ and $t_{z}\rightarrow \infty$ and $s_{z}\rightarrow \infty$, and where

\begin{eqnarray}
q^{2}_{B} = -\frac{(1-r)^2}{r}m^2_{B}-\frac{s_{x}^2}{r}\equiv -Q^{2}_{B}\,,\quad
{q^{2}_{B}}_{\mid s_{x}=0} &=& -\frac{(1-r)^2}{r}m^2_{B}\,.
   \label{limiteqn}
   \end{eqnarray}

We found that:

\begin{equation}\label{CRMain2eqn}
F_2^{B}(q^{2}_{B})=\frac{m^{2}_{B}}{m^{2}_{\Omega^{-}}}\, F_2^{\Omega^{-}}(q^{2}_{\Omega^{-}}),
\end{equation}

\begin{equation}\label{CRMain2eqnA}
F_1^{B}(q^{2}_{B})= F_1^{\Omega^{-}}(q^{2}_{\Omega^{-}}).
\end{equation}

{\sloppy
\noindent  Clearly, if one knows $F_1^{\Omega^{-}}(q^{2}_{\Omega^{-}})$ for some range $0\geq q^{2}_{\Omega^{-}} \geq q^{2}_{K}$, then one knows the value of $r_{K}\geq r\geq 1$ and thus $q^{2}_{B}$ (from Eq.~(\ref{limiteqn})) for this same range and hence one can infer $F_1^{B}(q^{2}_{B})$ and $F_2^{B}(q^{2}_{B})$ from Eqs.~(\ref{CRMain2eqn}) and (\ref{CRMain2eqnA}). We illustrate this in Fig.~1 where $B$ is the $\Delta^{-}$ (or the $\Delta^{+}$---see Eq.~(\ref{su2eqn2}) below) and $F_1^{\Delta^{-}}(q^{2}_{\Delta^{-}})$ is predicted using lattice calculations from Ref.~\cite{Alexandrou:2010jv}
for the $\Omega^{-}$ electric charge form factor (dipole fit).
}

 To obtain the magnetic moments of the $U$-Spin $1$, $\frac{1}{2}$, and $0$ decuplet baryons, {\bf one must find a way to quantitatively connect} the decuplet $U$-Spin multiplets.  We proceed to do this by first defining $\braket{Bs^{\sigma},3/2}{j_{em}^{\mu}(0)}{Bt^{\sigma},3/2}\equiv \brak{B}$, $\braket{Bs^{\sigma},3/2}{j_{V}^{\mu}(0)}{Bt^{\sigma},3/2}\equiv \brakt{B}$, and $\braket{Bs^{\sigma},3/2}{j_{S}^{\mu}(0)}{Bt^{\sigma},3/2}\equiv \braks{B}$ so that  $\brak{B}=\brakt{B}+ \braks{B}$ where $B$ is now any decuplet baryon).  Second, we utilize the double ETCRs to relate the matrix elements $\brak{\Delta^{-}}$, $\brak{\Delta^{0}}$, $\brak{\Delta^{+}}$, and $\brak{\Delta^{++}}$ (a $U$-Spin singlet) to each other and to that of the $\Omega^{-}$.  We can---for example---use $\left[V_{K^{0}},j^{\mu}_{em}(0)\right]=0$ to obtain the magnetic moment of the $\Sigma^{*\,+}$ from that of the $\Delta^{+}$ ($U$-Spin doublet) and the magnetic moments of the $\Sigma^{*\,0}$ and $\Xi^{*^0}$ from that of the $\Delta^{0}$ ($U$-Spin triplet).

The double ETCRs $\detcr{j^{\mu}_{em}(0)}{V_{\pi^{+}}}{V_{\pi^{-}}}$ $=$ $\detcr{j^{\mu}_{em\;3}(0)}{V_{\pi^{+}}}{V_{\pi^{-}}}$ $=$ $2j^{\mu}_{em\;3}(0)$ \cite{Oneda:1979mp} sandwiched between the pair states $\bra{\Delta^{++}},\ket{\Delta^{++}}$, $\bra{\Delta^{+}},\ket{\Delta^{+}}$, $\bra{\Delta^{0}},\ket{\Delta^{0}}$, and $\bra{\Delta^{-}},\ket{\Delta^{-}}$ can be used to determine the $SU(2)$ parametrization of $j^{\mu}_{em}(0)$ for the $\Delta$ states in the infinite momentum frame.  This produces six equations:

\begin{subequations}
\label{su2eqn}
\begin{eqnarray}
\brak{\Delta^{++}}=\brak{\Delta^{-}}-2\brakt{\Delta^{-}}\,,\quad \brakt{\Delta^{++}}=-\brakt{\Delta^{-}}\\
  \brak{\Delta^{+}}=\brak{\Delta^{-}}-\frac{4}{3}\brakt{\Delta^{-}}\,, \quad 3\brakt{\Delta^{+}}=-\brakt{\Delta^{-}}\\
  \brak{\Delta^{0}}=\brak{\Delta^{-}}-\frac{2}{3}\brakt{\Delta^{-}}\,\,, \qquad 3\brakt{\Delta^{0}}=\brakt{\Delta^{-}}.
  \end{eqnarray}
\end{subequations}



{\sloppy
Third, the axial-vector matrix elements (in the infinite momentum frame) \cite{Oneda:1979mp}\linebreak[4]
$\braket{\Delta^{+},3/2}{A_{\pi^{-}}}{\Delta^{++},3/2}$ = $\braket{\Delta^{-},3/2}{A_{\pi^{-}}}{\Delta^{0},3/2}$
$\equiv -\sqrt{3/2}\;\tilde{g}$, and $\braket{\Delta^{0},3/2}{A_{\pi^{-}}}{\Delta^{+},3/2}$
$=-\sqrt{2}\;\tilde{g}$
} and the double ETCR $\detcr{j^{\mu}_{em}(0)}{A_{\pi^{+}}}{A_{\pi^{-}}}$ $=2j^{\mu}_{em\;3}(0)$ sandwiched between the same pair $\Delta$ states allow us to write the following four equations:

\begin{subequations}
\label{axialeqn}
\begin{eqnarray}
3\tilde{g}^{2} [\brak{\Delta^{-}}-\brak{\Delta^{0}}]=4 \brakt{\Delta^{-}}\,,\\
3\tilde{g}^{2}[7\brak{\Delta^{0}}-3\brak{\Delta^{-}}-4\brak{\Delta^{+}}]=4 \brakt{\Delta^{-}}\,,\\
3\tilde{g}^{2}[-7\brak{\Delta^{+}}+4\brak{\Delta^{0}}+3\brak{\Delta^{++}}]=4 \brakt{\Delta^{-}}\,,\\
3\tilde{g}^{2} [\brak{\Delta^{+}}-\brak{\Delta^{++}}]=4 \brakt{\Delta^{-}}\,.
\end{eqnarray}
\end{subequations}

Finally, Eqs.~(\ref{axialeqn}) in conjunction with Eqs.~(\ref{su2eqn}) imply in broken symmetry that:

\begin{eqnarray}
\tilde{g}^{2}=2, \quad
\brak{\Delta^{++}}=-2\brak{\Delta^{-}}, \quad
  \brak{\Delta^{+}}=-\brak{\Delta^{-}},\quad \mbox{and} \quad
  \brak{\Delta^{0}}=0\,.
     \label{su2eqn2}
\end{eqnarray}

Eq.~(\ref{su2eqn2}) \emph{effectively connects the $U$-Spin $1$, $\frac{1}{2}$, and $0$ decuplet baryon matrix elements to that of the $U$-Spin $=\frac{3}{2}$ \,$\Delta^{-}$ (and hence the $\Omega^{-}$  }) and with Eq.~(\ref{CRMain2eqn}) and Eq.~(\ref{CRMain2eqnA})---Eq.~(\ref{limiteqn}) is valid for \emph{all} $U$-Spin decuplet baryons---allow us to compute the magnetic moments of the $\Delta^{++}$, $\Delta^{+}$, and $\Delta^{0}$ and their (strangeness $S\neq 0$) $U$-Spin partners in terms of $\Omega^{-}$ magnetic moment data by using the ETCR $\left[V_{K^{0}},j^{\mu}_{em}(0)\right]=0$ which results in:

\begin{eqnarray}
\mu_{B}=-Q_{B}\left[ 1-  \left(  \frac{m_{B}^2}{m_{\Omega^{-}}^2}   \right) \left( \frac{m + m_{\Omega^{-}}(\mu_ {\Omega^{-}}/\mu_{N}) }{m}  \right)    \right] \left(\frac{m}{m_{B}}\right)\,\mu_{N}.
\label{Maineqn}
\end{eqnarray}

 Eq.~(\ref{Maineqn}) is the \emph{main result of this work and is valid for all of the ground-state $J^{P}=\frac{3}{2}^{+}$ baryon decuplet members}. As the values of $m_{\Delta}$ (all $\Delta$ charge states) (pole or Breit-Wigner) are not very well established, we assume $m_{\Delta}= 1.22\pm 0.01\;GeV/c^{2}$.
Experimentally, we have \cite{Nakamura:2010zzi}, $\mu_{\Omega^{-}}=(-2.02 \pm 0.05)\,\mu_{N}=\left[(-1+
  F_2^{\Omega^{-}}(0)
  ) (m/m_{\Omega^{-}})\right] \mu_{N}$ and $m_{\Omega^{-}}=1.6724\pm 0.0003\;GeV/c^{2}$.  We summarize our results for all of the ground-state baryon decuplet magnetic moments $\mu_{B}$ in Table~I.

\section{Conclusions}

We have--nonperturbatively--calculated the magnetic moments of all of the ground-state $J^{P}={3/2}^{+}$ \emph{physical decuplet} baryons without ascribing any specific form to their quark structure or intra-quark interactions or assuming a Lagrangian (effective or otherwise).  The Particle Data Group \cite{Nakamura:2010zzi} value of $\mu_{\Omega^{-}}$ along with other decuplet mass data was used as input except we took $m_{\Delta}= 1.22\pm 0.01\;GeV/c^{2}$ (all $\Delta$ charge states) as the values of $m_{\Delta}$ are not well-enough established \cite{Nakamura:2010zzi}.
In particular---utilizing Eq.~(\ref{Maineqn})---we obtained $\mu_{\Delta^{-}}= (-1.83 \pm 0.04) \,\mu_{N}$, $\mu_{\Delta^{+}}= (+1.83 \pm 0.04) \,\mu_{N}$, and $\mu_{\Delta^{++}}= (+3.67 \pm 0.07) \,\mu_{N}$ and $\mu_{\Delta^{0}}= (0 ) \,\mu_{N}$.  Our results for the magnetic moments (the $\Omega^{-}$ magnetic moment is input) of the ground-state decuplet baryons are summarized in Table~I along with a prediction in Fig.~1 for the $\Delta^{-}$ (and the $\Delta^{+}$) electric charge form factor as a function of $Q^{2}$ based upon $\Omega^{-}$ lattice calculated fit data \cite{Alexandrou:2010jv}. Similarly---with Eq.~(\ref{su2eqn2})---one may predict the electric charge form factor for the $\Delta^{++}$ as a function of $Q^{2}$ based upon $\Omega^{-}$ lattice calculated fit data.
{\bf{\emph{For all}}} of the ground-state $J^{P}={3/2}^{+}$ baryons $B$, we have demonstrated how the $F_1^{B}(q^{2}_{B})$ and $F_2^{B}(q^{2}_{B})$ form factors can be calculated in in terms of $\Omega^{-}$ data.
Future experimental measurements of the $\Omega^{-}$ magnetic moment and accessible form factors for $q^{2}_{\Omega^{-}}< 0$ will have great importance for viable theoretical models (especially lattice QCD models) of the structure of baryons. Knowledge of the behavior of the decuplet form factors (or corresponding multipole moments) is critical to our understanding of QCD---standard model, enhanced standard model, lattice gauge models, superstring models, or entirely new models--- since these models must be capable of yielding already known results at low or medium energy.  Eqs. ~(\ref{CRMain2eqn}), ~(\ref{CRMain2eqnA}), ~(\ref{limiteqn}), and ~(\ref{su2eqn2}) explicitly demonstrate that the electromagnetic charge form factors of the decuplet baryons are very closely related to each other and that their magnetic dipole form factors are also very closely related to each other.  This may aid experimental and theoretical ground-state decuplet baryon magnetic moment analyses in the future.

\begin{table}[h]
\caption{\label{table1}ground-state baryon decuplet magnetic moment $\mu_{B}$ in units of $\mu_{N}$.}
\begin{ruledtabular}
\begin{tabular}{@{}lccc@{}}
Baryon B&  This research\;\footnotemark[1] & Particle Data Group;\footnotemark[2] &Lattice QCD\;\footnotemark[3]  \\
       &     &     &  \\
\hline
$\Delta^{++}$\hphantom{00} & \hphantom{0}$+3.67 \pm 0.07$ & $+5.6\pm 1.9$   & ---  \\
$\Delta^{+}$\hphantom{00} & \hphantom{0}$+1.83 \pm 0.04$ & $+2.7\pm 3.6$   & ---  \\
$\Delta^{0}$\hphantom{00} & \hphantom{0}$0 \pm 0$ & ---   & ---  \\
$\Delta^{-}$\hphantom{00} & \hphantom{0}$-1.83\pm 0.04$ & ---   &  $-1.85\pm 0.06$ \\
$\Sigma^{*\,+}$\hphantom{00} & \hphantom{0}$+1.89\pm 0.04$  & ---    & ---  \\
$\Sigma^{*\,0}$\hphantom{00} & \hphantom{0}$0\pm 0$  & ---    & ---  \\
$\Sigma^{*\,-}$\hphantom{00} & \hphantom{0}$-1.89\pm 0.04$  & ---    & ---  \\
$\Xi^{*\,0}$\hphantom{0} & \hphantom{0}$0\pm 0$ & ---    & --- \\
$\Xi^{*\,-}$\hphantom{0} & \hphantom{0}$-1.95\pm 0.05$ & ---    & --- \\
$\Omega^{-}$\hphantom{0} & \hphantom{0}$-2.02\pm 0.05$   & $-2.02\pm 0.05$    & $-1.93\pm 0.08$  \\
\end{tabular}
\end{ruledtabular}
\footnotetext[1]{$\mu_{\Omega^{-}}$ is input. $m_{\Delta}= 1.22\pm 0.01\,GeV/c^{2}$ is assumed for all $\Delta$ charge states. $\mu_{\Omega^{-}}$ and other baryon masses are from the Particle Data Group \cite{Nakamura:2010zzi}.  Statistical propagation of errors used in calculations. }
\footnotetext[2]{$\Delta^{++}$ estimate from Ref.~\cite{Nakamura:2010zzi}. $\Delta^{+}$ error (quadrature calculated) from Ref.~\cite{Nakamura:2010zzi} (see original Ref.~\cite{Kotulla:2003pm}). }
\footnotetext[3]{Lattice result from Ref.~\cite{Aubin:2009qp}.}
\end{table}

\newpage

\bibliographystyle{apsrev}

\begin{figure}[bp]
\includegraphics[scale=1,keepaspectratio=true,height=7in,width=6.5in]{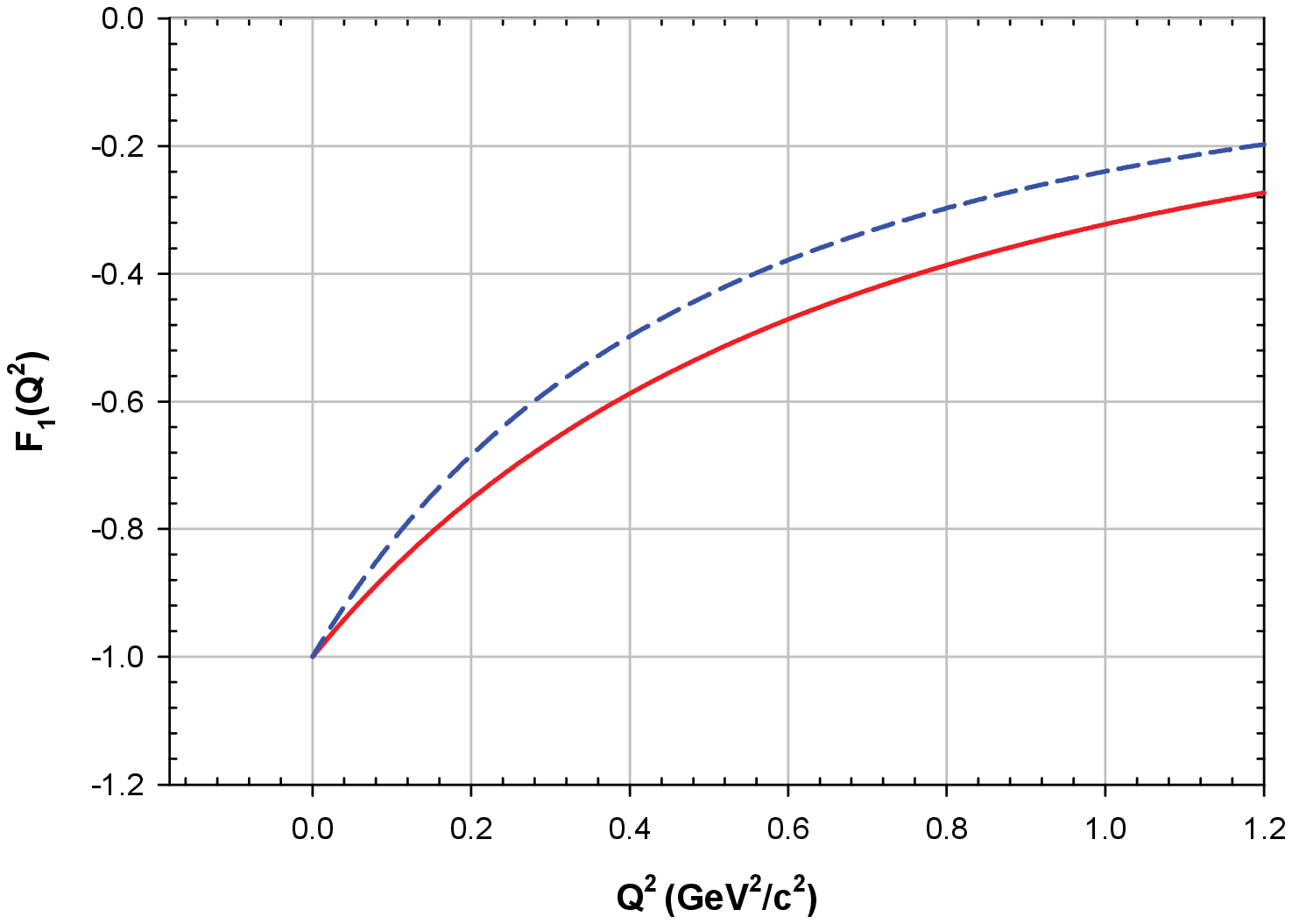}
\caption{The Solid curve is a dipole
fit [$(-1)(1+Q^{2}/\Lambda_{E_{0}}^{2})^{-2}$] with $\Lambda_{E_{0}}=1.146\;GeV/c$ to lattice calculations
 for the $\Omega^{-}$ electric charge form factor taken from Fig.~3 and Table III of Ref.\protect\cite{Alexandrou:2010jv}.
 The Dashed curve is the $\Delta^{-}$ electric charge form factor calculated using Eq.~(\ref{CRMain2eqnA}) and
Eq.~(\ref{limiteqn}) and the above $\Omega^{-}$ lattice dipole fit. The $\Delta^{+}$ $F_{1}(Q^2)$ electric charge form factor as a function of $Q^{2}$ is just (-1) times that of the Dashed curve according to Eq.~(\ref{su2eqn2}) and the assumption that $m_{\Delta}= 1.22\pm 0.01\;GeV/c^{2}$ for all $\Delta$ charge states.
}
\label{fig1}
\end{figure}

\end{document}